\documentstyle[12pt]{article}
\begin{document}
\vspace*{0.25in}
\newcounter{numeris}
\begin{center}
{\LARGE Rydberg Atoms Ionisation by Microwave Field and Electromagnetic
Pulses}\\[1.5\normalbaselineskip]
{\Large B. Kaulakys and G. Vilutis}\\[1.5\baselineskip]
{\small Institute of Theoretical Physics and Astronomy,\\
A. Go\v stauto 12, 2600 Vilnius, Lithuania}\\[2.5\baselineskip]
\end{center}

\begin{center}
{\small
\parbox{4.8in}{{\bf Abstract.} A simple theory of the Rydberg atoms ionisation
by
electromagnetic pulses and microwave field is presented. The analysis is based
on the scale transformation which reduces the number of parameters and reveals
the functional dependencies of the processes. It is shown that the observed
ionisation of Rydberg atoms by subpicosecond electromagnetic pulses scale
classically. The threshold electric field required to ionise a Rydberg state
may be simply evaluated in the photonic basis approach for the quantum
dynamics or from the multiphoton ionisation theory.}}
\end{center}
\vskip 2\baselineskip
\centerline{\large\bf INTRODUCTION}
\vskip \baselineskip
Highly excited atoms in electromagnetic fields serve as a simple important
examples of the strongly driven by an external driving field non-linear
systems with the stochastic behaviour and provide an unique opportunity to
explore the quantum manifestations of the classical chaos and the
quantum-classical correspondence for chaotic systems. That is why great
attention has been devoted to the experimental and theoretical investigations
of the dynamics of the Rydberg electron in the strong microwave field and to
ionisation peculiarities of the Rydberg states (1-4).

Recently the ultrashort, half-cycle electromagnetic pulses with central
frequencies around 1 $THz$ have been created in photoconducting switches and
used for ionisation of Rydberg atoms (5). One peculiarity of this process is
that the duration of the electromagnetic pulse is short compared to the
internal time, the Kepler orbital time $T_k=2\pi n_0^3$, of the atom under
study. This results to the specific features in scaling of the threshold
field required to ionise a Rydberg state. The classical theory explains the
half-cycle pulses (HCP) ionisation only qualitatively (5-7) while quantum
corrections in the numerical calculations (6-7) are negligible.

Here we show that (a) the observations (5) {\it consist with the general
classical scaling relations} for the hydrogenic atom in a microwave field (8),
(b) the same scaling and close threshold field results from the multiphoton
ionisation theory (1), and (c) it may be simply evaluated in the photonic
basis approach for the quantum dynamics.
\vskip 2\baselineskip
\centerline{\large\bf SCALING RELATIONS}
\vskip \baselineskip
The analysis of the classical and quantum dynamics of the hydrogenic atom in
a microwave field may be simplified introducing the scale transformation
which reduces the number of parameters and reveals the functional dependencies
of the processes (8-10).

The Hamiltonian in atomic units a.u. ($e=\hbar=m=1$) for the hydrogenic
atom in a microwave field of frequency $\omega$ and field strength $F$ is
$$
H={\bf p}^2/2-1/r+zF\sin(\omega\tilde t+\varphi)\eqno(1)
$$
where {\bf r} and {\bf p} are the position and momentum of the electron.
Measuring the time of the action of
the microwave field on the hydrogenic atom in the field periods one can
introduce the scale transformation (8):
$$
t=\omega\tilde t,~~~{\bf r}_s=\omega^{2/3}{\bf r},~~~{\bf p}_s={\bf p}/
\omega^{1/3},~~~F_s=F/\omega^{4/3},~~~H_s=H/\omega^{2/3}\eqno(2)
$$
where the scaled Hamiltonian is
$$
H_s={\bf p_s}^2/2-1/r_s+z_sF_s\sin(t+\varphi)\eqno(3)
$$
The scaled time-dependent Schr\" odinger equation can be expressed as
$$
i\omega^{1/3}{{\partial\Psi}\over{\partial t}}=(H_s^0+V_s)\Psi\eqno(4)
$$
$$
H_s^0={\omega^{2/3}\over 2}\left[-{1\over r_s^2}{\partial\over\partial r_s}
\left(r_s^2{\partial\over\partial r_s}\right)+{l(l+1)\over r_s^2}\right]-
{1\over{r_s}}
$$
$$
V_s=z_sF_s\sin(t+\varphi)\eqno(5)
$$
The scaled energy spectrum of the unperturbed hydrogen atom is
$$
E_s=-1/2\omega^{2/3}n^{2}=-1/2s^{2/3}\eqno(6)
$$
where $s=\omega/(-2E)^{3/2}$ is the relative field frequency: the ratio
of the microwave frequency $\omega$ to the Kepler orbital frequency of the
electron $\Omega=(-2E)^{3/2}$.

Expressions (3) shows that the classical motion of the electron with the
definite initial conditions depends only on the scaled field strength and,
on the contrary, the scaled threshold field strength for the onset of
classical chaos or for the ionisation of the Rydberg atom $F_s^{th}$ is a
function of the initial scaled energy or initial relative field frequency
$s_0=\omega n_0^3$, i.e. $F_s^{th}=f(s_0)$. The concrete form of the function
$f(s_0)$ depends on the ionisation mechanism, which is different for low,
intermediant and high relative frequencies $s_0$.
Namely, the static field $(s_0\to 0)$ ionisation threshold is $F_s^{st}\simeq
0.130/s_0^{4/3}$, while the threshold for onset of classical chaos in high
frequency $(s_0\gg 1)$ microwave field is at $F_s^{MW}\simeq 1/{49}s^{5/3}$,
Refs.
(1-4,8-10). From the finding, Ref. (5), of the threshold field for HCP
ionisation,
$F^{HCP}\simeq 0.3/n_0^2\tau_{HCP}^{2/3}$ if $s_0\ge 1$, we have the scaled
threshold field $F_s^{HCP}\simeq 0.14/s_0^{2/3}$. A simple classical model
explains the observed scaling but the theoretical and experimental ionisation
thresholds differ by a factor 2.5, Refs. (5-7, 10). So, {\it all the three
mechanisms
of classical ionisation are in the framework of the general classical scaling},
Ref. (8): the scaled threshold field is a function of the relative field
frequency, but the concrete form of the function depends on the ionisation
mechanism.

According to equations (4) and (5) the motion of the quantum hydrogen atom
in a monochromatic field is governed in addition to the scaled field by
the scaled Planck constant $\hbar=\omega^{1/3}$ (in a.u.).
\vskip 2\baselineskip
\centerline{\large\bf QUANTUM IONISATION THEORY}
\vskip \baselineskip
The direct first-order quantum ionisation of Rydberg atom by electromagnetic
pulses is relatively weak and exhibits no threshold field dependence since it
is proportional to $F^2$. Therefore, we consider the multistep ionisation
processes. For simplification of the problem one can introduce the photonic
basis and calculate quantum transitions in the model system with an equidistant
energy spectrum, Ref. (8). Using the known expressions for the dipole matrix
elements between excited atomis states one can evaluate the ionisation
probability by means of Prenyakov and Urnov's model (11) (see also, Refs.
(8-10)):
$$
P_i\simeq J_{N_i}^2(K)\simeq (2\pi N_i)^{-1}(eK/2N_i)^{2N_i},~~~K\simeq
\pi F_s/2\omega^{1/3}.\eqno(7)
$$
Here $J_N(K)$ is the Bessel function, $e\simeq 2.718\dots$, $N_i\simeq
1/2n_0^2\omega$
is the number of photons required for ionisation and the effective dipole
matrix elements for transitions between the photonic states were used. Thus,
the ionisation is appreciable if $eK/2N_i\simeq 1$, or $F_s^{Phb}\simeq
2/e\pi s_0^{2/3}$, which is close to the observations, Ref. (5). On the other
hand, the rate of the multiphoton ionisation is
$\omega_i\propto (7.05n_0^2F/\omega^{2/3})^{2N_i}$, Ref. (1). Therefore, the
threshold field for multiphoton ionisation is $F_s^{MPh}\simeq
1/7.05s_0^{2/3}$ which is {\it in agreement with the observed, Ref. (5),
scaling and absolute experimental results}.
\newpage
\centerline{\large\bf CONCLUSIONS}
\vskip \baselineskip
Scale transformations for the dynamics of the hydrogenic atom in an
electromagnetic field are wery useful and enables one to maximally simplify
the analyses of the dynamic processes and reduce the number of parameters of
the problem. The classical motion depends only on the scaled field strength
while the quantum dynamics depends, in addition, on the scaled Planck constant.

Simple analytical approach based on the photonic basis yields to the correct
description of the quantum dynamics for the hydrogen atom in a microwave field
including the quantum suppression of chaotic diffusion effect and enables one
to clarify relations between the different quantum-classical correspondence
conditions.

The observed scaling in the ionisation of Rydberg atoms by subpicosecond
half-cycle electromagnetic pulses consist with the general classical scaling
for the hydrogenic atom in a microwave field. The same scaling and close
threshold fields for ionisation result from the multiphoton ionisation
theory and may be simply evaluated in the photonic basis approach. The
observed peculiarity of the threshold field scaling results from the shortness
of the electromagnetic pulses.
\vskip 1.5\baselineskip
\centerline{\large\bf ACKNOWLEDGMENTS}
\vskip \baselineskip
The research described in this publication was made possible in part by Grant
No. LAA000 from the International Science Foundation.
\vskip 1.5\baselineskip
\centerline{\large\bf REFERENCES}
\vskip \baselineskip
\small
\begin{list}{\arabic{numeris}.\hfil}%
{\usecounter{numeris}\labelwidth=0.20in\leftmargin=0.25in\labelsep=0.05in%
\itemsep=0in\parsep=0in}
\item Delone N. B., Krainov V. P., and Shepelyansky D. L., {\it Usp. Fiz. Nauk}
{\bf 140}, 355-392 (1983) [{\it Sov. Phys. - Usp.} {\bf 26}, 551 (1983)].

\item Casati G., {\it et al.}, {\it IEEE J. Quantum
Electronic} {\bf 24}, 1420-1444 (1988).

\item Jensen R. V., Suskind S. M., and Sanders M. M., {\it Phys. Rep.} {\bf
201},
1-56 (1991).

\item Koch P. M., Atomic and Molecular Physics Experiments in Quantum Chaology,
in {\it Lecture Notes in Physics,} V. {\bf 411,} New York: Springer-Verlag,
1992, pp. 167-224.

\item Jones R. R., You D., and Bucksbaum P. H., {\it Phys. Rev. Lett.} {\bf
70},
1236-1239 (1993).

\item Reinhold C. O., Melles M., and Burgd\" orfer J., {\it Phys. Rev. Lett.}
{\bf 70}, 4026 (1993).

\item Reinhold C. O. {\it et al.}, {\it J. Phys. B.: At. Mol. Opt. Phys.} {\bf
26},
L659-L664 (1993).

\item Kaulakys B. {\it et al.}, {\it Phys. Lett.} {\bf 159A}, 261-265 (1991).

\item Kaulakys B., {\it Acta Phys. Polonia B} {\bf 23}, 313-316 (1992).

\item Kaulakys B., Gontis V., and G. Vilutis, {\it Lithuan. Phys. J.} {\bf 33},
354-357 (1993).

\item Presnyakov L. P. and Urnov A. M., {\it J. Phys. B} {\bf 3}, 1267-1271
(1970).
\end{list}
\end{document}